\documentclass[reprint,aps,pra,twocolumn,superscriptaddress,longbibliography]{revtex4-1}

\usepackage[utf8]{inputenc}
\usepackage{lmodern}
\usepackage{graphicx}
\usepackage{bm}
\usepackage{amsmath,amssymb,amsfonts}
\usepackage{physics}
\usepackage{xcolor}
\usepackage{bbold}
\usepackage{hyperref}

\hypersetup{
  colorlinks=true,
  linkcolor=blue,
  urlcolor=blue,
  citecolor=blue
}

\begin{document}

\title{Fundamental error bound for entanglement generation\\ between interacting Rydberg atoms}

\author{Georgios Doultsinos}
\affiliation{Institute of Electronic Structure and Laser and Center for Quantum Science and Technologies, FORTH, 70013 Heraklion, Crete, Greece}
\affiliation{Department of Physics, University of Crete, Heraklion, Greece}

\author{Antonis Delakouras}
\affiliation{Institute of Electronic Structure and Laser and Center for Quantum Science and Technologies, FORTH, 70013 Heraklion, Crete, Greece}
\affiliation{Department of Physics, University of Crete, Heraklion, Greece}

\author{David Petrosyan}
\affiliation{Institute of Electronic Structure and Laser and Center for Quantum Science and Technologies, FORTH, 70013 Heraklion, Crete, Greece}
\affiliation{A. Alikhanyan National Laboratory (YerPhI), 0036 Yerevan, Armenia}

\begin{abstract}
We analytically derive the lower error bound for the preparation of any maximally entangled state of two atoms involving Rydberg-state interactions. 
This fundamental bound represents the minimum achievable error $E \geq ( 1 + \pi/2 ) \Gamma/B$ due to spontaneous decay $\Gamma$ of the Rydberg states and their finite interaction strength $B$, assuming that all other technical errors can be eliminated. 
Using quantum optimal control methods, we identify laser pulses for preparing a maximally entangled state of a pair of atomic qubits with an error only $1\%$ above the derived fundamental bound. 
\end{abstract}

\date{\today}
\maketitle

\section{Introduction}

Neutral atoms in arrays of reconfigurable microtraps \cite{barredo2016atom,endres2016atom,barredo2018synthetic} represent one of the leading platforms for physical implementation of quantum computing \cite{xia2015, Levine2019Parallel, graham2019, graham2022, evered2023high, evered2026high, ma2023high, Tsai2025, MunizPRXQ.2025}. 
In these systems, the main mechanism to realize quantum gates or produce entanglement between atomic qubits relies on laser-excitation of the atoms to Rydberg states \cite{Saffman2010InformationRydberg}, i.e., states with high principal quantum number $n$. 
Rydberg states possess large (transition or induced) dipole moments $\wp \propto n^2$ \cite{Gallagher1994Rydberg}, which enables them to interact strongly over mesoscopic distances via the van der Waals or dipole-dipole interactions with strength $B\propto n^{11}/x^6$ or $\propto n^4/x^3$, respectively, depending on the selected state and the interatomic distance $x$. 
Moreover, Rydberg states are relatively stable and, in cryogenic environments, spontaneously decay with small but non-negligible rates $\Gamma \propto n^{-3}$. 
This requires quantum gates and entangling operations, such as preparation of a Bell state, to be performed in short times $T \ll \tau = 1/\Gamma$, while minimizing the population of the decaying states, in order to achieve high fidelities. 

The speed of these operations is limited, in the blockade regime $B \gg \Omega$, by the amplitude $\Omega$ of the laser pulse driving the atoms between the ground and Rydberg states, and the time-optimal pulses have been determined in Ref. \cite{Jandura2022} and employed in most experiments ever since \cite{evered2023high, ma2023high, Tsai2025, MunizPRXQ.2025}. 
But ultimately the error $E=\Gamma T_r = \eta \Gamma/B$ is determined by the average time $T_r$ that the atoms spend in the decaying Rydberg states, and is proportional to the ratio $\Gamma/B$ and scaling coefficient $\eta$ that depends on the laser pulses, assuming that all the technical errors are negligible or can in principle be eliminated (e.g., atomic motion, leakage to other Rydberg states, laser phase and amplitude noise, external electric and magnetic field fluctuations, etc.). 
Hence, identifying pulses that minimize $T_r$, and thereby $\eta$, is highly desirable and important, as has been highlighted in several works \cite{Saffman2010InformationRydberg, Saffman2016, Poole2025}, but, to the best of our knowledge, never satisfactorily addressed.

Here we show that the decay error $E$ for creating any maximally entangled state of two atoms, or qubits, is bounded by $E \geq \eta_{\min} \Gamma/B$ where $\eta_{\min} = \left( 1 + \pi/2 \right)$. 
Furthermore, we demonstrate that this bound is tight: 
We identify laser pulses that generate a maximally entangled (Bell) state with error only about $\sim 1\%$ above this theoretical limit. 
This result refines the previous estimate \cite{Wesenberg2007}, which (after correcting numerical inaccuracies, see Appendix~\ref{app:WeakBound}) yielded a looser lower bound of $E \gtrsim 1.05\, \Gamma/B$.

\section{Lower bound for decay error} 
\label{sec:bound}
We consider two atoms, $A$ and $B$, each possessing $d \geq 2$ internal electronic states, one of which is an excited Rydberg state $\ket{r}$. 
The total Hamiltonian of the system $\mathcal{H} = \mathcal{H}_\mathrm{int} + \mathcal{H}_\mathrm{loc}$ consists of the interaction and local terms.
We assume dispersive interaction given by ($\hbar = 1$)
\begin{equation}
\mathcal{H}_\mathrm{int} = B\,\ket{rr}\bra{rr} .
\label{Eq:vdW_interaction}
\end{equation} 
The local term $\mathcal{H}_\mathrm{loc} = \mathcal{H}_A \otimes\mathbb{1}_B + \mathbb{1}_A\otimes\mathcal{H}_B$ describes single-atom laser driving between the Rydberg state and the remaining stable, non-interacting electronic states. The precise form of $\mathcal{H}_\mathrm{loc}$ is left unspecified at this stage.

At any time $t$, we can use the Schmidt decomposition to express the state of the system as 
\begin{equation}
    \ket{\psi(t)} = \sum_{i=1}^dc_i(t)\, \ket{u_i(t)}\ket{v_i(t)} , \label{eq:Schmidt}
\end{equation}
where $c_i(t) \ge 0$ are the time dependent Schmidt coefficients satisfying $\sum_i c_i^2(t) = 1$, while $\{\ket{u_i}\}, \{\ket{v_i}\}$ are orthonormal bases in the Hilbert spaces of atoms $A$ and $B$, respectively. 
Our goal is to transform an initial product state $\ket{\psi(0)}\equiv \ket{\psi_\mathrm{in}}$ with $c_1(0) = 1$ and $c_{i>1}(0)=0$ into a maximally-entangled (Bell) state $\ket{\psi(T)}\equiv\ket{\psi_\mathrm{f}} $ with $c_1(T) = c_2(T) = 1/\sqrt{2}$ and $c_{i>2}(T)=0$, with the highest possible fidelity $F = \big|\bra{\psi_\mathrm{f}}\mathcal{U}(T)\ket{\psi_{\mathrm{in}}}\big|^2 \simeq 1$,
where $\mathcal{U}(t) = \mathcal{T}\exp\{-i\int_0^t\mathcal{H}(t')\,dt'\}$ is the unitary time-evolution operator and $T$ is the process duration.

For $T \ll\tau =1/\Gamma$, decay of the Rydberg states with rate $\Gamma$ can be accounted for via the non-Hermitian contribution $\tilde{\mathcal{H}} = \mathcal{H} - i\frac{\Gamma}{2}\,\Pi_r,$ where $\Pi_r = \sum_{j = A,B} \ket{r}_j\bra{r}$ is a projector onto the Rydberg states.
Since entanglement generation necessarily involves populating the interacting Rydberg states, the resulting decay error is given by
\begin{equation}
E \simeq \Gamma T_r \equiv \Gamma \int_0^T P_r(t)\, dt,
\end{equation}
where $T_r$ is the average time that the atoms spend in the Rydberg states and $P_r(t) = \bra{\psi(t)} \Pi_r \ket{\psi(t)}$ is the instantaneous Rydberg-state population of the system in state $\ket{\psi(t)} = \mathcal{U}(t) \ket{\psi_{\mathrm{in}}}$. 
Hence, our goal is to perform the transformation $\ket{\psi_\mathrm{in}} \to \ket{\psi_\mathrm{f}}$ while minimizing the accumulated decay error $E$. 
The smallest possible $E$ will constitute the fundamental lower bound of error for entanglement generation.

To calculate the bound, we follow an analysis similar to that of Ref.~\cite{Vidal2001} and consider the so-called min-entropy $S = -\log_2 c_1^2$, where $c_1\equiv c_\mathrm{max}$ is the largest Schmidt coefficient. 
The min-entropy is a convenient measure of the distance of a state from a maximally entangled state.
Clearly, $S(\psi_\mathrm{in}) = 0$ for the initial state, and $S(\psi_\mathrm{f}) = 1$ for the final state 
\footnote{Note that a state with $S=1$ is not necessarily a Bell state, except when $d=2$.
Nevertheless, determining the minimum decay error for attaining states with $S=1$ provides a bound that also applies to the special case of Bell-state preparation.}. 
Substituting the Schmidt decomposition (\ref{eq:Schmidt}) into the Schrödinger equation $i \frac{d}{dt} \ket{\psi} = \mathcal{H} \ket{\psi}$
and omitting the explicit time dependence of variables to simplify notation, we find that $\dot{c}_1 = \sum_{i=1}^d c_i\,\mathrm{Im}\bra{u_1 v_1}\mathcal{H}\ket{u_i v_i}$ (see Appendix~\ref{app:WeakBound}).
The time derivative of the min-entropy is then
\begin{equation}
    \dot{S}(\psi) = -\frac{2}{\ln{2}}\frac{\sum_{i=1}^d c_i\,\mathrm{Im}\bra{u_1 v_1}\mathcal{H}_\mathrm{int}\ket{u_i v_i}}{c_1} ,
\end{equation}
where we replaced $\mathcal{H}$ with $\mathcal{H}_\mathrm{int}$, since for unitary evolution the min-entropy can change only due to interaction-induced dynamics.

We can now determine the bound for the average time $T_r$ that the atoms spend in the Rydberg states as follows.
We can write  
\begin{equation}
    \label{eq:changeVar}
    T_r \equiv \int_0^TP_r(t)\,dt = \int_0^T \frac{P_r(\psi(t))}{|\dot{S}(\psi(t))|}\,|\dot{S}(\psi(t))|\,dt .
\end{equation}
Assuming the min-entropy monotonically increases from $S=0 \to 1$ (see Appendix~\ref{app:coArea} for a more general case), we can change the variable $t \to s = S(\psi_s)$ obtaining
\begin{equation}
    \label{eq:changeVar2}
    T_r = \int_0^1 \frac{P_r(\psi_s)}{|\dot{S}(\psi_s)|}\,ds\geq \int_0^1 G(s)\,ds ,
\end{equation}
where 
\begin{equation}
    G(s) \equiv \min_{\psi_s} 
    \frac{P_r(\psi_s)}{|\dot{S}(\psi_s)|} ,
    \label{eq:GofS}
\end{equation}
and the minimum is taken over all states $\ket{\psi_s}$ satisfying $S(\psi_s) = s$, or, equivalently, all states with $c_1 = 2^{-s/2}$.

A closed-form expression for $G$ can be obtained analytically (see Appendix~\ref{app:varCalc}) and it leads to $c_{i>2}=0$, thus implying $c_2 =\sqrt{1-2^{-s}}$.
This yields $G = \frac{\ln 2}{B}\, \frac{c_1}{c_2}(c_1+c_2)^2$, or, in terms of the min-entropy,
\begin{equation}
    G(s) = \frac{\ln 2 }{B}\: \frac{(1+\sqrt{2^s-1})^2}{2^{s}\sqrt{2^s-1}}.
    \label{eq:GofSanalytic}
\end{equation}
Integrating $G(s)$ over the interval $s\in[0,1]$ yields $T_r \geq (1+\pi/2)/B$ and therefore
\begin{equation}
E \geq \eta_{\min} \frac{\Gamma}{B} , \quad \eta_{\min} = 1+\frac{\pi}{2}.
\label{eq:fundLim}
\end{equation}
From the same analysis, we can also determine the time duration $T$ of the transformation as 
\begin{equation}
T = \int_0^1 \frac{ds}{\dot{S}}, \label{eq:Tint}
\end{equation}
where, for the $\ket{\psi_s}$ that minimizes the right-hand side of Eq.~(\ref{eq:GofS}), we have 
\begin{equation}
    \dot{S} = B\,\frac{2}{\ln 2}\,\frac{2^s - 1}{(1 + \sqrt{2^s - 1})^2}.
    \label{eq:dotSanalytic}
\end{equation}
Remarkably, the integral in Eq.~(\ref{eq:Tint}) diverges as $s \to 0$, implying that the optimal evolution that saturates inequality~\eqref{eq:fundLim} can be attained only in the limit of $BT \to \infty$. 
Nevertheless, below we identify finite-duration pulses for the preparation of a maximally-entangled (Bell) state with error that approaches the fundamental bound \eqref{eq:fundLim} to within $\sim1\%$ deviation.

\section{Laser pulse via GRAPE optimization}
\label{sec:GRAPE}
\begin{figure}[t]
\includegraphics[width=1\linewidth]{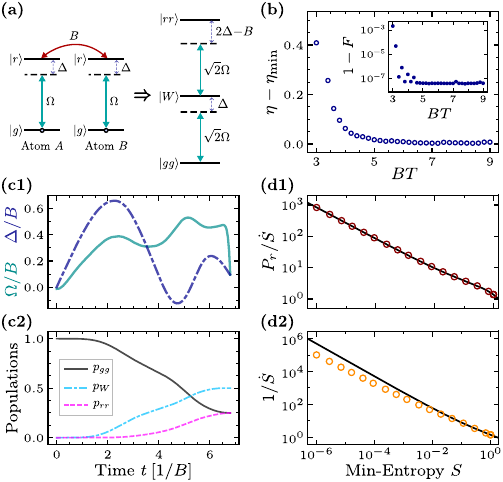}
    \caption{(a) Level scheme of two two-level atoms with the transition between the stable ground $\ket{g}$ and decaying Rydberg $\ket{r}$ states driven by a laser with Rabi frequency $\Omega$ and detuning $\Delta$. 
    Atoms in state $\ket{r}$ interact dispersively with strength $B$.
    The system is equivalent to a three-level ladder system with states $\ket{gg}, \, \ket{W}=(\ket{gr} + \ket{rg})/\sqrt{2}, \, \ket{rr}$. 
    (b)~Scaling coefficient $\eta$ (with $\eta_{\min} =1+\pi/2$) vs laser pulse duration $T$, as obtained by GRAPE optimization of preparation of target state $\ket{\psi_{\mathrm{f}}}$. 
    Inset shows the preparation fidelity $F$ under unitary evolution.
    (c)~Temporal profiles of the Rabi frequency $\Omega(t)$ (solid cyan line) and detuning $\Delta(t)$ (blue dashed line) for an optimal pulse of duration $BT=6.8$ and area $\Theta\simeq2.31$, and the corresponding dynamics of populations of states $\ket{gg}, \, \ket{W}, \, \ket{rr}$. 
    For this pulse, the corresponding infidelity is $1-F \simeq 4 \times 10^{-8}$ and scaling coefficient is $\eta \simeq 2.575$. 
    (d)~Comparison between the theoretical bounds (solid lines) and the solution identified by GRAPE optimization (open circles). 
    The GRAPE solution $\ket{\psi}$ closely follows the theoretical minimum $P_r(\psi)/\dot{S}(\psi)\simeq G ( S(\psi))$ of Eq.~(\ref{eq:GofS}), despite the finite pulse duration $T$. In contrast, $1/\dot{S}(\psi)$ diverges for $S \to 0$ more slowly than the theoretical result, which requires $BT\to\infty$ as per Eq.~(\ref{eq:Tint}).}
    \label{fig:PulsesPop}
\end{figure}

The minimal system to exhibit entanglement corresponds to a pair of two-level atoms with the internal states \(\ket{g}\) and \(\ket{r}\) that can serve as qubit basis states $\{ \ket{0}, \ket{1} \}$.  
For alkali or alkaline-earth atoms commonly used in experiments \cite{Levine2019Parallel, graham2019, graham2022, evered2023high, ma2023high, Tsai2025, MunizPRXQ.2025}, \(\ket{g}\) corresponds to a long-lived lower electronic state, while \(\ket{r}\) is a highly-excited Rydberg state.
These states are coupled by laser field(s) with the time-dependent Rabi frequency $\Omega(t)$ and detuning $\Delta(t)$, see Fig.~\ref{fig:PulsesPop}(a). 
The lasers act symmetrically on both atoms $A$ and $B$, while we do not assume the blockade regime, i.e., both $\Omega(t)$ and $\Delta(t)$ are arbitrary (but smooth) functions of time $t$.  
The local Hamiltonian is thus
\begin{equation}
\mathcal{H}_\mathrm{loc} = \tfrac{1}{2} \Omega(t) \sum_{j} \big(\sigma_{rg}^{(j)} + \sigma_{gr}^{(j)}\big)
- \Delta(t) \sum_{j} \sigma_{rr}^{(j)},
\end{equation}
where \(\sigma_{\mu \nu}^{(j)} = \ket{\mu}_j\bra{\nu}\) is the transition ($\mu\neq \nu$) or projection ($\mu = \nu$) operator for atom $j=A,B$.

A naive approach to entangle the two atoms initially in state $\ket{\psi_\mathrm{in}} = \ket{gg}$ is to 
apply a very strong ($\Omega \gg B$) and short resonant $\pi/2$-pulse to both atoms to prepare them in state $\tfrac{1}{2}(\ket{g} + \ket{r})(\ket{g} + \ket{r})$. 
Then, with $\Omega = 0$, one can wait for time $T = \pi/B$, during which the system accumulates the dynamical interaction phase $\phi_{\mathrm{int}} = BT =\pi$ and evolves into the maximally entangled state $\ket{\psi_\mathrm{f}} = \tfrac{1}{2}(\ket{gg} + \ket{gr} + \ket{rg} - \ket{rr})$ 
\footnote{This state is equivalent to the Bell state $(\ket{gg} + \ket{rr})/\sqrt{2}$, up to a single atom ($A$ or $B$) rotation $\ket{g,r} \to (\ket{g} \pm \ket{r})/\sqrt{2}$ by a resonant $\pi/2$-pulse.}. 
This protocol corresponds to the fastest possible way to generate entanglement between the two atoms, but it involves the Rydberg-state population $P_r(t) = 1\:\forall\, t$, hence $T_r=T$, leading to the decay error $E \gtrsim \pi\, \Gamma/B$ that exceeds the fundamental bound \eqref{eq:fundLim} by a factor of $\sim 1.22$.

We therefore consider dynamical laser driving of the system initially in state $\ket{\psi_\mathrm{in}} = \ket{gg}$ and set the target state $\ket{\psi_\mathrm{f}} = \tfrac{1}{2}\big(\ket{gg} + \ket{rg} + \ket{gr} - \ket{rr}\big)$. 
Since the Hamiltonian is symmetric under exchange of the atoms, the dynamics is confined to the symmetric subspace of the two-atom Hilbert space, $H_{\mathrm{sym}} = \mathrm{span}\{\ket{gg},\, \ket{W},\, \ket{rr}\}$, where $\ket{W} = \tfrac{1}{\sqrt{2}}(\ket{gr} + \ket{rg})$, and we neglect single-atom dephasing that could couple the bright state $\ket{W}$ to the dark state $\ket{D}=\tfrac{1}{\sqrt{2}}(\ket{gr}-\ket{rg})$.
Within this subspace [see Fig.~\ref{fig:PulsesPop}(a)], the system reduces to a three-level ladder system governed by the effective Hamiltonian 
\begin{equation}
\mathcal{H}' =
\begin{pmatrix}
    0 & \tfrac{1}{\sqrt{2}}\Omega & 0 \\
    \tfrac{1}{\sqrt{2}}\Omega & -\Delta & \tfrac{1}{\sqrt{2}}\Omega\\
    0 & \tfrac{1}{\sqrt{2}}\Omega & B -2\Delta 
\end{pmatrix}.
\end{equation}  
Our task is to prepare the state 
$\ket{\psi_\mathrm{f}} = \tfrac{1}{2}\big( \ket{gg} + \sqrt{2}\ket{W} - \ket{rr} \big)$ using a laser pulse with time-dependent $\Omega (t),\Delta(t)$, whereas the interaction strength $B$ is fixed, while minimizing the Rydberg-state population $P_r$ which is the expectation value of operator \(\Pi_r' = \mathrm{diag}(0, 1, 2)\).  
This problem does not admit a trivial solution.

To determine the optimal pulses, we employ the gradient ascent pulse engineering (GRAPE) method \cite{Khaneja2005}.  
For the numerical optimization, we set the total evolution time $T$ and divide it into $N \gg 1$ equal time steps $t_n=n\, \mathrm{d}t$ with $\mathrm{d}t=T/N$. We then optimize simultaneously over the discrete values of the Rabi frequency  $\Omega_n = \Omega(t_n)$ and the detuning $\Delta_n = \Delta(t_n)$ for $n = 0, \dots, N$ 
\footnote{Note that since $B$ is finite, the GRAPE optimization must find both the pulse amplitude $\Omega(t)$ and phase $\varphi(t)$ [or detuning $\Delta(t) = d \varphi(t)/dt$], unlike the case of $B \to \infty$ where the time-optimal pulse is the one with maximum $\Omega = \mathrm{const}$ for all $t\in [0,T]$ and only the phase $\varphi(t)$ is a function of time \cite{Jandura2022}.}.  
The parameter space thus consists of $2(N + 1)$ variables, and we evaluate the gradients of the fidelity $\delta F / \delta \Omega_n$ and $\delta F / \delta \Delta_n$, see Appendix~\ref{app:Grads}.

In addition to maximizing the preparation fidelity $F$ under the unitary evolution, the solution must minimize $T_r $ and thereby the decay error $E=\Gamma T_r$. We therefore define the cost functional
\begin{equation}
J(\gamma) = F - \gamma T_r, \label{eq:OptCost}
\end{equation}
where \(\gamma > 0\) controls the strength of the penalty.  
During the optimization, \(\gamma\) is gradually reduced from \(10^{-3} B\) to zero. 
This strategy, commonly used in constrained optimization \cite{Nocedal2006}, helps the algorithm to first explore solutions with low Rydberg population and not necessarily optimal fidelity \(F=1\). 
Once the optimization has converged to a good region of the parameter space, the penalty $\gamma$ is progressively relaxed to focus on maximizing the fidelity.
The gradients $\delta T_r / \delta \Omega_n$ and $\delta T_r / \delta \Delta_n$ are also computed (see Appendix~\ref{app:Grads}) and combined with those of the fidelity to construct the overall gradient of the cost function. 
The total gradients are then used within the BFGS (Broyden--Fletcher--Goldfarb--Shannon) algorithm \cite{Nocedal2006} to identify optimal solutions.

In Fig.~\ref{fig:PulsesPop}(b) we show the scaling coefficient $\eta$ for the smallest decay error $E=\eta \Gamma/B$ for the preparation of target state $\ket{\psi_{\mathrm{f}}}$ with optimal pulses of various durations $T$. Clearly $\eta \to \eta_{\min}$ for sufficiently long pulses $BT \gtrsim 5$. 
In the inset we also show the best fidelity $F$ of preparation of $\ket{\psi_{\mathrm{f}}}$ obtained for $\gamma \to 0$ in Eq.~(\ref{eq:OptCost}). With decreasing $T$, the infidelity $1-F$ increases, diverging for $TB \to \pi$, and the optimizer fails to identify pulses yielding $F=1$.

In Fig.~\ref{fig:PulsesPop}(c) we show the Rabi frequency $\Omega(t)$ and detuning $\Delta(t)$ of an optimal pulse for the preparation of the target state $\ket{\psi_\mathrm{f}}$ and the corresponding populations of states $\ket{gg}, \ket{W}, \ket{rr}$.  
The total duration of this pulse is $T = 6.8/B$ and its area is $\Theta = \int_0^T |\Omega(t)| dt \simeq 2.31$. 
The achieved infidelity is $1 - F < 10^{-7}$ and the average time spent in the Rydberg-state is $T_r \simeq 2.575/B$, corresponding to a decay error $E = \Gamma T_r$ only $\sim 1\%$ above the fundamental bound (\ref{eq:fundLim}).  
In Fig.~\ref{fig:PulsesPop}(d) we demonstrate that the GRAPE solution $\ket{\psi}$ indeed minimizes $P_r(\psi)/\dot{S}(\psi) \simeq G(S(\psi))$, as obtained from the theoretical analysis in Eq.~(\ref{eq:GofSanalytic}). 
At the same time, the corresponding $1/\dot{S}(\psi)$ diverges with $S \to 0$ slower than the theoretical result for optimal preparation.
This is because the GRAPE solution corresponds to finite duration $T$ of the process, whereas the optimal solution requires infinite $T$, as per Eqs.~(\ref{eq:Tint}) and~\eqref{eq:dotSanalytic}.
   
\section{Discussion}
To summarize, using general arguments, we have derived the fundamental lower bound \eqref{eq:fundLim} for the error $E$ during the preparation of maximally entangled states of a pair of atoms interacting via decaying Rydberg states. 
Our derivation of the error bound is mathematically rigorous and can be applied to other similar systems for quantum information processing and provide useful fidelity bounds in the presence of decay or other relaxation processes.
This bound also applies to the average error of any two-qubit Rydberg gate that can maximally entangle four orthogonal input product states, such as the CZ gate, i.e. 
$\min \bar{E} = \Gamma\int_0^T \bar{P}_r(t)\,dt \geq \eta_{\min} \Gamma/B$. Here 
$\bar{P}_r = \tfrac{1}{4}\sum_{q_A q_B}\bra{q_A q_B}\mathcal{U}^\dagger(t) \, \Pi_r \,\mathcal{U}(t)\ket{q_A q_B}$ is the instantaneous Rydberg state population averaged over the four input states $\ket{q_A q_B}$ subject to the 
unitary evolution $\mathcal{U}(t)$. 

Next, using GRAPE optimization, we have identified smoothly-varying laser pulses of finite duration $T\gtrsim 5/B$ that, starting with the two-atom state $\ket{gg}$, prepare a maximally entangled (Bell-like) state $\ket{\psi_{\mathrm{f}}}$ with error $E \simeq 2.575\, \Gamma/B$, very close to the lower bound \eqref{eq:fundLim}, which implies that this bound is tight and realistically achievable. 
The corresponding unitary transformation $\mathcal{U}(t)$ is, by construction, a perfect entangler with information content $c = (c_1, c_2, c_3) \simeq (1.7, 0.31, 0.31)$ \cite{Zhang2003}. 
It is, however, not a special perfect entangler \cite{Rezakhani2004}, i.e., it does not simultaneously maximally entangle the four orthogonal input product states of two atoms,
while the resulting average decay error is large, $\bar{E} = 6.8\Gamma/B$, since $\bar{P}_r =1 \, \forall \, t \in [0,T]$ and $T=6.8/B$. 

We were not able to determine a pulse that implements a special perfect entangling gate with an average decay error close to the fundamental bound $\bar{E} \simeq (1+\pi/2) \Gamma/B$  \footnote{A maximally entangled (Bell) state between two atoms can be converted into a gate using additional internal degrees of freedom of the atoms, local unitary transformations, followed by measurements and teleportation: R. Jozsa, An introduction to measurement based quantum computation, NATO Science Series, III: Computer and Systems Sciences. Quantum Information Processing-From Theory to Experiment, \textbf{199}, 137-158 (2005); \href{https://doi.org/10.48550/arXiv.quant-ph/0508124}{arXiv:quant-ph/0508124}.}.  
Instead, for various $d \ge 2$, the optimizer identified pulses implementing quantum gates with the average errors $\bar{E} \gtrsim \pi \Gamma/ B$. 
This result is straightforward for a pair of two-level atoms ($d=2$) with 
$\ket{g}\equiv\ket{0}$ and $\ket{r}\equiv\ket{1}$:
The best strategy is to simply wait for time $T = \pi/B$ until the interacting state $\ket{rr} \equiv \ket{11}$ accumulates a dynamical $\pi$-phase, while all the other states $\ket{gg} \equiv \ket{00}, \ket{gr} \equiv \ket{01}, \ket{rg} \equiv \ket{10}$ remain unchanged, corresponding to the CZ gate. 
Since the Rydberg state population averaged over all the inputs is $\bar{P}_r (t) = 1 \, \forall \, t$, the resulting average decay error is $\bar{E} = \pi \Gamma/ B$. 
Similar logic applies to a pair of three-level atoms ($d=3$) with two stable ground-state sublevels representing the qubit states, $\ket{g_0} = \ket{0}$ and $\ket{g_1} = \ket{1}$.  
In this case, the optimal pulse sequence consists of a strong ($\Omega \gg B$) resonant $\pi$-pulse that induces the transition $\ket{1} \rightarrow \ket{r}$, followed by a waiting period of $T = \pi/B$, and concluded by a second strong $\pi$-pulse that de-excites the atoms $\ket{r} \rightarrow \ket{1}$.
We note that the previous best theoretical attempt \cite{Poole2025} to identify laser pulses for realizing a CZ Rydberg gate and entangling two atoms led to the decay error $\bar{E} \simeq 3.4 \Gamma/B$. 
For the gates with highest overall fidelity achieved in the experiments so far \cite{evered2026high} the decay error is $\bar{E} \simeq 200 \Gamma/B$ and it contributes to about a third of the total gate error $\bar{E}_{\mathrm{total}} \simeq 1.5 \times 10^{-3}$.
Hence, finding experimentally-relevant laser pulses to implement Rydberg quantum gates with minimal error remains a challenge, especially for the strong interaction (blockade) regime $B\gg \max \Omega$. 

\begin{figure}[t]
\includegraphics[width=1\linewidth]{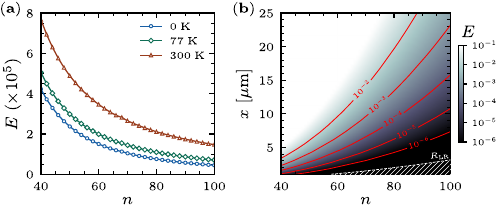}
    \caption{(a) Lower bound of the decay error $E=\eta_{\mathrm{min}} \Gamma/B$ for the preparation of entanglement of a pair of $^{87}$Rb atoms excited to the Rydberg state $\ket{r}=|nS_{1/2}\rangle$ with the principle quantum number $n$ for different temperatures $0,77,300\:$K.
    For each $n$, we calculate the decay rates $\Gamma$ and van der Waals coefficient $C_6$ and place the atoms at the distance $x=3 R_{\mathrm{LR}}$, where $R_{\mathrm{LR}}$ is the LeRoy radius, obtaining thereby the interaction strengths $B=C_6/x^6$. 
    (b) Minimum error $E$ for different quantum numbers $n$ for a pair of atoms separated by distances $x$, with the $C_6$ and $\Gamma$ (temperature 0K) as in (a). Contours bound regions in the $x-n$ parameters space with the error $E < 10^{-2}, 10^{-3}, \ldots , 10^{-6}$. White dashed line indicates the LeRoy distance $x=R_{\mathrm{LR}}$ below which the results do not apply. 
    Atomic data are calculated using the ARC Python library~\cite{ARClib2017Pritchard}. }
    \label{fig:Evsnx}
\end{figure}

It is instructive to consider the implications of our results for neutral atom quantum computers. 
To be specific, we consider alkali atoms, such as $^{87}$Rb, widely used in experiments \cite{Levine2019Parallel, graham2019, graham2022, evered2023high, evered2026high}. 
In Fig. \ref{fig:Evsnx}(a), we show the minimal decay error $E=\eta_{\mathrm{min}} \Gamma/B$ vs the principal quantum number $n$ of the Rydberg state $\ket{r}=|nS_{1/2}\rangle$ of atoms at small separation $x=3 R_{\mathrm{LR}}$ and for different temperatures. 
For $n=40-60$, the minimal error $E < 10^{-4}$ can be achieved even at room temperatures using laser pulses driving the ground-Rydberg transition $\ket{g}\to\ket{r}$ with large but experimentally realistic Rabi frequencies $\Omega/2\pi \simeq 20-40\:$MHz.  
For higher Rydberg states, $n\leq 100$, and cryogenic environment, the minimal error can be even smaller, $E \sim 10^{-5}$. 
In Fig. \ref{fig:Evsnx}(b) we show the minimal error for different $n$ and interatomic distances $x$ that determine the van der Waals interaction strength $B=C_6/x^6$. 
The important observation here is that entanglement or quantum gates with sufficiently low error bound $E \leq 10^{-3}$ can be obtained even for large interatomic separation $x \gtrsim 10\:\mu$m by going to the higher Rydberg states with $n \gtrsim 80$ \cite{Pecorari2025,Poole2025}.
Very similar results are obtained for $^{133}$Cs.

\paragraph*{Note added --} 
Our method was recently adopted to derive fundamental error bounds for quantum gates employing resonant dipole-dipole exchange (F\"orster) interaction $V$ between pairs of Rydberg states of atoms \cite{Norrell2026}. 
It was found that, for rank-one coupling (a single laser driving the ground-Rydberg transition of each atom) the minimal gate error $E \geq \eta_{\min} \Gamma/V$ is the same, $\eta_{\min} = 1+\pi/2$, as found here for dispersive interactions $B$ of Eq.~(\ref{Eq:vdW_interaction}). 
But for a more involved rank-two coupling scheme (two lasers coupling each Rydberg state to the corresponding ground state of each atom) the minimal gate error can be smaller, $\eta_{\min} = \pi/2$.

\section*{Acknowledgments} 
This work was supported by the EU programme HORIZON-CL4-2021-DIGITAL-EMERGING-01-30 via the Project EuRyQa (Grant No. 101070144).

\section*{Data availability} 
The data that support the findings of this article are openly available at \cite{Doultsinos2025Bounds_Data}.

\bibliography{refs}

\newpage

\appendix

\section{Technical derivations}

\subsection{Application of the coarea formula}\label{app:coArea}

During the evolution, the min-entropy $S(t)$ does not necessarily increase monotonically from $S=0 \to 1$.
But even for non-monotonic $S(t)$, we can use the co-area formula for one-dimensional integrals to derive the inequality \eqref{eq:changeVar2}.

Let \( g(x): [a, b]\to\mathbb{R} \) be a continuously differentiable function (not necessarily monotone) such that \( g'(x) \) is non-zero except possibly at isolated points. Also let \( f(x):[a,b]\to[0, \infty) \) be integrable. Then the co-area formula states that
\begin{equation}
    \int_a^b f(x)\,|g'(x)|\,dx = \int_\mathbb{R} \left(\sum_{x\in g^{-1}(y)} f(x)\right) dy,
\end{equation}
where the set \( g^{-1}(y) = \{x \mid g(x)=y\} \) is the inverse image (preimage) of \( y \). If \( g(x) \) is monotone, the co-area formula reduces to the usual change of variables in one-dimensional integrals.

We can now apply this formula for \( P_r(t)=P_r(\psi(t)) \) and the min-entropy \( S(t) = S(\psi(t)) \) as 
\begin{align}
    \int_0^T P_r(\psi(t))\,dt 
    &= \int_0^T \frac{P_r(\psi(t))}{|\dot{S}(\psi(t))|}\,|\dot{S}(\psi(t))|\,dt \nonumber\\
    &= \int_0^1 \left(\sum_{\psi(t)\in S^{-1}(s)} \frac{P_r(\psi(t))}{|\dot{S}(\psi(t))|}\right) ds \nonumber\\
    &\ge \int_0^1 \min_{\psi_s} \frac{P_r(\psi_s)}{|\dot{S}(\psi_s)|}\,ds,
\end{align}
where \( \psi_s \) is any state satisfying \( S(\psi_s) = s \), and in the last line, we used the fact that the sum over the preimages \( \psi(t)\in S^{-1}(s) \) is bounded from below by its smallest term.

\subsection{Analytic expression for $G(s)$} \label{app:varCalc}

Our goal here is to obtain an analytic expression for the bound $G(s) = \min_{\psi_s}\frac{P_r(\psi_s)}{|\dot{S}(\psi_s)|}$. 
We first express the states $\ket{\psi_s}$ in Schmidt form as $\ket{\psi_s} = \sum_i c_i \ket{u_i}\ket{v_i}$, where $c_1 \equiv c_\mathrm{max}= 2^{-s/2}$, and define $w_i = \frac{1}{2}\left(|\langle r | u_i \rangle|^2 + |\langle r | v_i \rangle|^2\right)$, which satisfy $0 \leq w_i \leq 1$ and $\sum_iw_i =1$. 
The entanglement rate can be bounded from above as
\begin{align}
    \lvert\dot{S}(\psi_s)\rvert &\leq \frac{2B}{\ln 2 \, c_1}\sum_{i>1} c_i\lvert\mathrm{Im}\{\bra{u_1v_1}\ket{rr}\bra{rr}\ket{u_iv_i}\}\rvert \nonumber\\
    &\leq \frac{2B}{\ln 2 \, c_1}\sum_{i>1} c_i|\bra{u_1v_1}\ket{rr}\bra{rr}\ket{u_iv_i}| \nonumber\\
    &\leq \frac{2B}{\ln 2 \, c_1}\sum_{i>1} c_i w_1 w_i.
\end{align}
In the first line, we used the triangle inequality. In the second line, we applied the inequality for complex numbers $|\mathrm{Im}\,\zeta| \leq |\zeta|$, which in this case is saturated when $\arg \langle rr|u_iv_i\rangle - \arg\langle rr|u_1v_1\rangle = \pm \pi/2$. In the third line, we used the inequality $w_i \geq |\langle r|u_i\rangle \langle r|v_i\rangle|$, which is saturated if and only if $|\langle r|u_i\rangle| = |\langle r|v_i\rangle|$.
Moreover, the average population in the Rydberg states can be expressed as $P_r = 2\sum_i c_i^2 w_i$. Combining the two expressions, we obtain
\begin{align}
    \frac{P_r}{|\dot{S}|} &\geq \frac{\ln2\,c_1}{B}
    \frac{\sum_i c_i^2 w_i}{\sum_{i>1} c_i w_1 w_i}\nonumber\\
    &= \frac{\ln2 \, c_1}{B} \frac{c_1^2 w_1 + \sum_{i>1} c_i^2 w_i}{w_1 \sum_{i>1} c_i w_i}.
\end{align}

To simplify the notation and the algebra, we define the variables $x = c_1^2 =2^{-s}$, $y = w_1$, where $1/2 \leq x \leq 1$ since $s \in [0,1]$, and also introduce $a_i^2 = c_i^2/(1-x)$ and $b_i = w_i/(1-y)$, for which the normalization $\sum_{i>1} a_i^2 = \sum_{i>1} b_i = 1$ holds. Hence, the expression becomes
\begin{equation}
    \frac{P_r}{|\dot{S}|} \geq \frac{\ln2 }{B} \sqrt{\frac{x}{1-x}}\frac{xy + (1-x)(1-y)\sum_{i>1} a_i^2 b_i}{y(1-y)\sum_{i>1}a_i b_i}.
\end{equation}
Using the Cauchy–Schwarz inequality, we have 
\begin{equation}
    \left(\sum_{i>1}a_i b_i\right)^2 \leq \sum_{i>1}a_i^2 b_i \sum_{j>1} b_j = \sum_{i>1} a_i^2 b_i,
\end{equation}
where we used the normalization of $b_j$. 
We define $z = \sum_{i>1} a_i b_i$ which satisfies $0 \leq z \leq 1$ due to the normalization of both $a_i$ and $b_i$. Thus, the expression becomes 
\begin{equation}
    \frac{P_r}{|\dot{S}|} \geq \frac{\ln2 }{B} \sqrt{\frac{x}{1-x}}\frac{xy + (1-x)(1-y)z^2}{z}.
\end{equation}

Since the variable $x = 2^{-s}$ is fixed, our goal is to minimize the function
\begin{equation}
    f(x;\,y, z) = \frac{xy + (1-x)(1-y)z^2}{y(1-y)z} 
\end{equation}
with respect to $y$ and $z$. 
To find the critical points of this function, we solve $\partial_y f = \partial_z f = 0$ and obtain the following system of equations:
\begin{subequations}
    \begin{align}
        y &= \frac{z\sqrt{1-x}}{\sqrt{x} + z\sqrt{1-x}},\\
        z^2 &= \frac{xy}{(1-x)(1-y)}.
    \end{align}
\end{subequations}
Substituting one equation into the other, we find that the critical point of $f$ is
\begin{equation}
    z_* = \sqrt{\frac{x}{1-x}}, \qquad   y_* = \frac{1}{2}.
\end{equation}
However, for $x > 1/2$ it follows that $z_* > 1$ which is false. This indicates that the critical point lies outside the domain of function $f$. 
In other words, the minimum must occur on the boundary of the domain, at $z = 1$. This corresponds to $a_2 = b_2 = 1$ and $a_i = b_i = 0$ for $i > 2$. 
Equivalently we can choose any other index $i_*$ so that $a_{i_*} = b_{i_*} = 1$ and $a_i=b_i=0$ for $i\neq i_*$. Hence, 
\begin{equation}
    f_\mathrm{min}(x) =  \min_{y}f(x;\,y, 1) = (\sqrt{x} + \sqrt{1 - x})^2 .
\end{equation}
Upon substituting back $x = 2^{-s}$, we obtain
\begin{align}
    G(s) &= \frac{\ln2}{B} \sqrt{\frac{2^{-s}}{1-2^{-s}}} f_\mathrm{min}(2^{-s}) \nonumber \\
    &= \frac{\ln 2 }{B}\: \frac{(1+\sqrt{2^s-1})^2}{2^{s}\sqrt{2^s-1}}.
\end{align} 
This expression has also been verified numerically by minimizing $P_r(\psi_s)/|\dot{S}(\psi_s)|$ over all states $\ket{\psi_s}$ for which $S(\psi_s) = s$, for different values of $s \in [0, 1]$.

\subsection{Gradients for GRAPE optimization} \label{app:Grads}

The gradients of the fidelity with respect to the control parameters, for the discretization $t_n=n\, \mathrm{d}t$ with $\mathrm{d}t = T/N$ and $n=1, \dots, N$, are given by
\begin{subequations}
    \begin{align}
        \frac{\delta F}{\delta \Omega_n} &= 2\mathrm{d}t \,\mathrm{Im}\{ \bra{\chi_n} \mathcal{V}_\Omega \ket{\psi_n}\bra{\psi_n}\ket{\chi_n}\},\\
        \frac{\delta F}{\delta \Delta_n} &= 2\mathrm{d}t \,\mathrm{Im}\{ \bra{\chi_n} \mathcal{V}_\Delta \ket{\psi_n}\bra{\psi_n}\ket{\chi_n}\}.
    \end{align}
\end{subequations}
Here, $\mathcal{V}_\Omega = \tfrac{1}{2}\sum_j (\sigma_{rg}^{(j)} + \sigma_{gr}^{(j)})$ and $\mathcal{V}_\Delta = -\sum_j \sigma_{rr}^{(j)}$, with $j = A, B$.  
The states $\ket{\psi_n} = \ket{\psi(t_n)}$ are obtained from the forward Schrödinger evolution $i\frac{d}{dt}\ket{\psi(t)} = \mathcal{H}(t)\ket{\psi(t)}$, with $ \ket{\psi(0)} = \ket{\psi_\mathrm{in}}.$
The states $\ket{\chi_n} = \ket{\chi(t_n)}$ are obtained from solving $i\frac{d}{dt}\ket{\chi(t)} = \mathcal{H}(t)\ket{\chi(t)}$ with the boundary condition $\displaystyle{\ket{\chi(T)}} = \ket{\psi_\mathrm{f}}.$

For the gradients of the average time spent at the Rydberg states $T_r =\int_0^T P_r(t) \,dt$, we introduce auxiliary states $\ket{\xi_n} = \ket{\xi(t_n)}$ satisfying the inhomogeneous Schrödinger equation $i\frac{d}{dt}\ket{\xi(t)} = \mathcal{H}(t)\ket{\xi(t)} + \Pi_r \ket{\psi(t)}$ with $ \ket{\xi(T)} = 0$, which can be derived using first-order perturbation theory.
The corresponding gradients are then given by
\begin{subequations}
    \begin{align}
        \frac{\delta T_r}{\delta \Omega_n} &= 2\mathrm{d}t \,\mathrm{Re}\{ \bra{\xi_n} \mathcal{V}_\Omega \ket{\psi_n}\},\\
        \frac{\delta T_r}{\delta \Delta_n} &= 2\mathrm{d}t \,\mathrm{Re}\{ \bra{\xi_n} \mathcal{V}_\Delta \ket{\psi_n}\} .
    \end{align}
\end{subequations}



\section{Detailed derivation of the decay error in Ref. \cite{Wesenberg2007}}
  \label{app:WeakBound}
Here we revisit the derivation of the optimal scaling for the decay error when generating a unit of entanglement, as originally reported in \cite{Wesenberg2007}. There, it was found that $E \geq \eta_\mathrm{min} \Gamma / B$ with $\eta_\mathrm{min} \simeq 2.09$. By carefully following the logic of the original derivation, we find that the corresponding scaling coefficient is actually smaller by a factor of two, i.e., $ \eta_\mathrm{min} \simeq 1.05$.

We consider a bipartite system with Hilbert space $H = H_A \otimes H_B$ and write its Hamiltonian in the general form $\mathcal{H} = \mathcal{H}_A \otimes \mathbb{1}_B + \mathbb{1}_A \otimes \mathcal{H}_B + \mathcal{H}_{\mathrm{int}}$. 
The amount of entanglement is quantified by the von Neumann entropy of either subsystem, i.e., $S_\mathrm{vN} = S_{\mathrm{vN}}(\rho_A) = S_{\mathrm{vN}}(\rho_B)$, where $\rho_{A,B} = \mathrm{Tr}_{B,A}(\ket{\psi}\bra{\psi})$ denotes the reduced density matrix of the corresponding subsystem.

We express the total wavefunction $\ket{\psi}$ in its Schmidt decomposition as
\begin{equation}
    \ket{\psi} = \sum^d_{i=1} c_i \ket{u_i v_i}, \label{Eq:Schmidt_decomp}
\end{equation}
where $c_i \geq 0$ and $\sum_i c_i^2 = 1$, and $\ket{u_i}, \ket{v_i}$ form orthonormal bases in $H_A$ and $H_B$, respectively. Defining $p_i \equiv c_i^2$, the von Neumann entropy reads $S_\mathrm{vN} = -\sum_i p_i \log_2(p_i)$. The rate of entanglement generation is then
\begin{align}
    \dot{S}_\mathrm{vN} &=-\sum_i \frac{dp_i}{dt} \log_2(p_i) - \sum_i \frac{1}{\ln 2} \frac{dp_i}{dt} \nonumber\\
    &= -\sum_i \frac{dp_i}{dt} \log_2(p_i),
\end{align}
where in the second line we used that $\sum_i \frac{dp_i}{dt} = 0$. 

Substituting \eqref{Eq:Schmidt_decomp} into the Schrödinger equation $i \frac{d}{dt} \ket{\psi} = \mathcal{H} \ket{\psi}$ ($\hbar=1$), we obtain
\begin{align}
    i \left(\sum_i \frac{dc_i}{dt} \ket{u_i v_i} + \sum_i c_i \frac{d}{dt} \ket{u_i v_i} \right) = \sum_i c_i \mathcal{H} \ket{u_i v_i}.
\end{align}
Multiplying from the left by $\bra{u_j v_j}$ and using $\langle u_i \vert u_j \rangle = \langle v_i \vert v_j \rangle = \delta_{ij}$ gives
\begin{equation}
    \frac{dc_j}{dt} + \sum_i c_i \bra{u_j v_j} \frac{d}{dt} \ket{u_i v_i} = -i \sum_i c_i \bra{u_j v_j} \mathcal{H} \ket{u_i v_i}, \label{Eq:Schmidt_real_imag}
\end{equation}
where
\begin{align}
    \sum_i c_i \bra{u_j v_j} \frac{d}{dt} \ket{u_i v_i} &= \sum_i c_i \big[\bra{u_j v_j} \ket{\dot{u}_i v_i} + \bra{u_j v_j} \ket{u_i \dot{v}_i} \big] \nonumber\\
    &= c_j \big[\bra{u_j} \ket{\dot{u}_j} + \bra{v_j} \ket{\dot{v}_j} \big] \in \mathrm{Im}, \nonumber
\end{align}
since $\frac{d}{dt} \langle u_j \vert u_j \rangle = 0 \iff \langle u_j \vert \dot{u}_j \rangle = - \langle u_j \vert \dot{u}_j \rangle^*$ and similarly $\langle v_j \vert \dot{v}_j \rangle = - \langle v_j \vert \dot{v}_j \rangle^*$. Equating the real parts of both sides of eq. \eqref{Eq:Schmidt_real_imag} gives
\begin{equation}
    \frac{dc_i}{dt} = \sum_j c_j \, \mathrm{Im} \langle u_i v_i \rvert \mathcal{H} \rvert u_j v_j \rangle.
\end{equation}

Writing $\frac{dp_i}{dt} = 2 c_i \frac{dc_i}{dt}$, the rate of change of the von Neumann entropy can then be expressed as 
\begin{align}
    \dot{S}_\mathrm{vN} &=-2\sum_{ij}c_ic_j\log_2(c_i^2)\: \mathrm{Im}\bra{u_iv_i} \mathcal{H} \ket{u_jv_j} \nonumber\\
    &= \sum_{ij}c_ic_j\log_2\Big(\frac{1}{c_i^2}\Big)\: \mathrm{Im}\bra{u_iv_i} \mathcal{H} \ket{u_jv_j} \nonumber\\
    &+  \sum_{ij}c_jc_i\log_2\Big(\frac{1}{c_j^2}\Big)\:\mathrm{Im}\bra{u_iv_i} \mathcal{H} \ket{u_jv_j}^*, \nonumber 
\end{align}
where on the second and third lines we split the sum into two parts and interchanged indices $i\leftrightarrow j$ in the last. 
Using $\mathrm{Im}\bra{u_iv_i} \mathcal{H}\ket{u_jv_j}^*=-\mathrm{Im}\bra{u_iv_i} \mathcal{H}\ket{u_jv_j}$ and the properties of the logarithm, we obtain
\begin{equation}
    \dot{S}_\mathrm{vN} = 2\sum_{ij}c_ic_j\log_2\left(\frac{c_j}{c_i}\right) \mathrm{Im}\bra{u_iv_i}\mathcal{H}\ket{u_j v_j}. \label{Eq:Entanglement rate}
\end{equation}
Clearly, local terms do not contribute to generating entanglement between the two subsystems, and we may therefore replace $\mathcal{H}$ with $\mathcal{H}_\mathrm{int}$.

We now focus on the specific case of atoms interacting via $\mathcal{H}_{\text{int}} = B  \ket{r r}\bra{r r}$. Taking the absolute value of Eq.~\eqref{Eq:Entanglement rate} and using the triangular inequality $|\sum_{ij}\alpha_{ij}|\leq\sum_{ij}|\alpha_{ij}| \; \forall \;  \alpha_{ij}\in\mathbb{R}$, as well as $|\mathrm{Im} \,\zeta|\leq |\zeta| \; \forall \; \zeta \in \mathbb{C}$ gives
\begin{equation}
    \frac{|\dot{S}_\mathrm{vN}|}{B}\leq 2\sum_{ij}c_ic_j\left|\log_2\left(\frac{c_j}{c_i}\right)\right| \left|\langle u_iv_i \ket{rr}\right| \left|\langle rr|u_jv_j \rangle\right|.
\end{equation}
Employing now the inequality $\alpha\beta\leq \tfrac{1}{2}(\alpha^2 + \beta^2)$ for $(\alpha,\beta)=(\left|\langle v_{i} \ket{r}\right|,\left|\langle u_{i} \ket{r}\right|)$ and defining $w_i=\frac{1}{2}\left(|\bra{u_i}r\rangle|^2 + |\bra{v_i}r\rangle|^2\right)$, we have 
\begin{equation}
    \frac{|\dot{S}_\mathrm{vN}|}{B}\leq 2\sum_{ij}c_ic_j\left|\log_2\left(\frac{c_j}{c_i}\right)\right| w_iw_j. 
\end{equation}
Note here that $\sum_iw_i=1$, since $\{\ket{u_i}\},\{\ket{v_i}\}$ form orthogonal bases in the corresponding Hilbert spaces.
Parametrizing the Schmidt coefficients as $(c_i, c_j)=\sqrt{c_i^2+c_j^2}\left(\cos\theta_{ij}, \sin\theta_{ij}\right)$ gives 
\begin{align*}
     \frac{|\dot{S}_\mathrm{vN}|}{B} \leq & 2\sum_{ij}w_iw_j\left(c_i^2+c_j^2\right)\cos\theta_{ij}\sin\theta_{ij}\left|\log_2\left(\tan\theta_{ij}\right)\right|  \\
     =& 2\Big[\sum_ic_i^2w_i\sum_jw_j\cos\theta_{ij}\sin\theta_{ij}\left|\log_2\left(\tan\theta_{ij}\right)\right| \\
     &+\sum_ic_i^2w_i\sum_jw_j \sin\theta_{ij}\cos\theta_{ij}\left|\log_2\left(\cot\theta_{ij}\right)\right|\Big], \\
\end{align*}
where in the last line we exchanged indices $i\leftrightarrow j$ in the second sum and used $\sin{\theta_{ij}} = \cos{\theta_{ji}}$. Using now that $\left|\log_2\tan\theta_{ij}\right|=\left|\log_2\cot\theta_{ij}\right|$, we obtain
\begin{align}
    \frac{|\dot{S}_\mathrm{vN}|}{B} &\leq 4\sum_i c_i^2 w_i \sum_j f(\theta_{ij}) w_j, \label{Eq:Bound_f}
\end{align}
where we defined $f(\theta) = \sin\theta \cos\theta \left|\log_2(\tan\theta)\right|$, which satisfies $f(\theta) \lesssim 0.478$. 
We note that the factor $4$ appearing in Eq.~\eqref{Eq:Bound_f} differs from the factor $2$ reported in Ref.~\cite{Wesenberg2007}.

Replacing $f(\theta)$ with its maximum value and recalling that $\sum_jw_j=1$, the entanglement rate is bounded by $|\dot{S}_\mathrm{vN}|/B \lesssim 1.91\sum_ic_i^2w_i$. Moreover, the average population in the Rydberg state is given by 
\begin{align}
    P_r &= \bra{\psi}\left(\ket{r}\bra{r}\otimes\mathbb{1} + \mathbb{1}\otimes\ket{r}\bra{r}\right)\ket{\psi} \nonumber\\
    &= \sum_{ij}c_ic_j \left( \bra{u_i}r\rangle\bra{r}u_j\rangle\delta_{ij} + \delta_{ij}\bra{v_i}r\rangle\bra{r}v_j\rangle \right) \nonumber\\ 
    &= \sum_{i}c_i^2\left(|\bra{u_i}r\rangle|^2 + |\bra{v_i}r\rangle|^2\right)= 2 \sum_{i}c_i^2w_i , \nonumber\\
\end{align}
and therefore 
\begin{equation}
    |\dot{S}_\mathrm{vN}|\lesssim 0.956\: P_r B.
\end{equation}
Integrating the entanglement rate over the interval $[0,T]$ such that $S_\mathrm{vN}(0)=0$ and $S_\mathrm{vN}(T)=1$, i.e., we create a unit of entanglement, and multiplying by the decay rate $\Gamma$, we obtain
\begin{equation}
     E = \Gamma\int_0^T \! P_r dt \gtrsim 1.05 \frac{\Gamma}{B} .
\end{equation}

\end{document}